\documentclass{eas}
\usepackage{graphicx}
\usepackage{natbib}
\def\hho{H$_2$O}
\def\hhop{H$_2$O$^+$}
\def\pow#1#2{#1$\times$10$^{#2}$}
\def\msol{M$_{\odot}$}
\def\gtsim{{_>\atop{^\sim}}}

\begin{document}


\title{Herschel-HIFI observations of \hho\ in high-mass star-forming regions: first results}
\runningtitle{Van der Tak \etal: \hho\ in high-mass star-forming regions}
\author{Floris van der Tak}\address{SRON Netherlands Institute for Space Research, Landleven 12, 9747 AD Groningen, The Netherlands}
\author{Fabrice Herpin}\address{Universit\'e de Bordeaux, Laboratoire  d'Astrophysique, CNRS/INSU, UMR 5804, Floirac, France}
\author{Friedrich Wyrowski}\address{Max-Planck-Institut f\"ur Radioastronomie, Auf dem H\"ugel 69, 53121 Bonn, Germany}
\author{the high-mass WISH team}
%
\begin{abstract}
This paper reviews the first results of observations of \hho\ line emission with Herschel-HIFI towards high-mass star-forming regions, obtained within the WISH guaranteed time program. The data reveal three kinds of gas-phase \hho: `cloud water' in cold tenuous foreground clouds, `envelope water' in dense protostellar envelopes, and `outflow water' in protostellar outflows. The low \hho\ abundance (10$^{-10}$--10$^{-9}$) in foreground clouds and protostellar envelopes is due to rapid photodissociation and freeze-out on dust grains, respectively. The outflows show higher \hho\ abundances (10$^{-7}$--10$^{-6}$) due to grain mantle evaporation and (probably) neutral-neutral reactions.
\end{abstract}
\maketitle
\section{Introduction}

In normal galaxies like our own, most stars have about the mass of the Sun, and only $\sim$1\% of stars is more massive than 10\,\msol. Despite this rarity, high-mass stars are a major source of radiative and mechanical energy input to the interstellar medium, through ionizing UV radiation, strong stellar winds, and supernova explosions. High-mass stars are thus important for the evolution of their host galaxies, and form a link to starburst galaxies and the early Universe, where our view of star formation is dominated by the high-mass range.

The formation of high-mass stars is less well understood than the low-mass case, due to large distances, short timescales, and heavy extinction. Scenarios of monolithic accretion via a disk have had some success after suitable modification, in particular increased temperatures and stronger turbulence in the parent cloud. Alternatively, pre-stellar cores or protostellar envelopes may merge and/or accrete from the same reservoir. In any case, feedback is important due to the clustered nature of high-mass star formation.

While the angular resolution of ALMA will be essential to understand high-mass star formation on the `disk' scale, Herschel-HIFI will clarify the picture on the `envelope' and 'cloud' scales in at least three ways. First, HIFI gives full access to the \hho\ molecule, which is a key probe of interstellar physics and chemistry. Second, observations of hydride molecules are valueable probes of gas processing by radiation and shocks. Third, broad-band spectral surveys give a full view of the chemical composition of the gas, which is sensitive to parameters which are not directly observable, such as energetic (X-ray / cosmic-ray) irradiation and time. This review concentrates on \hho\ observations; the papers by Benz and Ceccarelli in this volume treat hydrides and spectral surveys, respectively.

The \hho\ molecule is a sensitive probe of physical conditions in interstellar gas clouds, in particular kinetic temperature and volume density. Being a major carrier of oxygen, the third most abundant element in the Universe, it influences the abundances of many molecular species. Previous space telescopes (ISO, Spitzer, SWAS, Odin) have observed \hho\ lines, but had limited spectral or angular resolution, or limited line coverage. The HIFI instrument on ESA's Herschel space observatory gives the first high-resolution view of the bulk of interstellar \hho, and \hho\ is the subject of a dedicated Guaranteed Time program, `Water in Star-forming regions with Herschel' (WISH; \citealt{ewine:wish}).
This paper describes the results of the first observations within the high-mass subprogram of WISH. For a precise overview of all planned observations in this subprogram, and the results for low-mass protostars, see the contributions by Herpin and Kristensen in this volume.

\section{Results}


The first observations were taken toward the DR21 region by \citet{vdTak:DR21}. Eighteen positions along a North-South strip were observed in the p-\hho\ $1_{11}$--$0_{00}$ and $^{13}$CO 10--9 lines. The \hho\ line profiles show broad absorption by the molecular outflow, narrow emission from the protostellar envelope, and narrow absorption by a foreground cloud. The envelope and the outflow are also seen in $^{13}$CO emission, while the foreground cloud is known from ground-based data. Abundances of \hho, estimated with radiative transfer programs, are (10$^{-10}$--10$^{-9}$) in the foreground cloud and the protostellar envelope, and a few 10$^{-7}$ in the outflow. Presumably photodissociation limits the \hho\ abundance in the foreground cloud, which has a low extinction, while freeze-out on dust grains limits the abundance in the dense envelope. In the outflow, grain mantle evaporation liberates \hho\ molecules into the gas phase.

These results are confirmed and extended by the multi-line study of \hho\ towards W3 IRS5 by \citet{chavarria:w3irs5}. The same basic three-component structure is seen in the line profiles of \hho, and in addition, the broad emission from the outflow was found to be very strong in the lines from the excited $J$=2 states. This result suggests a high temperature for the outflow ($\gtsim$200\,K), high enough that neutral-neutral reactions may enhance the gas-phase \hho\ abundance.
In addition, the spectra of this source show evidence for an enhanced \hho\ abundance in the warm inner part of the dense protostellar envelope, where temperatures are high enough for the thermal evaporation of icy grain mantles.

The low \hho\ abundances in high-mass protostellar envelopes are confirmed in the multi-source study by \citet{marseille:envelopes}. Using radiative transfer techniques, \hho\ abundances between \pow{5}{-10} and \pow{4}{-8} are derived; the spread in these values does not seem to be linked to physical properties of the sources. In addition, several foreground clouds are found in \hho\ $1_{11}$--$0_{00}$ absorption.

The relation between the \hho\ and \hhop\ molecules was studied by \citet{wyrowski:h2o+}. Of 10 sources observed, 9 show \hhop\ absorption, even when the \hho\ line appears in emission. While \hhop\ is most abundant in molecular outflows, it is also detectable in protostellar envelopes and foreground clouds. The \hhop/\hho\ abundance ratio is low in protostellar envelopes and higher in outflows and foreground clouds. Such a trend with volume density is consistent with the expectation that \hhop\ is mostly present in gas with a significant fraction of hydrogen in atomic form \citep{neufeld:hf}.

\section{Conclusions and outlook}

The first results from the high-mass WISH program show that around high-mass protostars, \hho\ is present in three distinct physical components: envelopes, outflows, and foreground clouds. The abundance of \hho\ is low in the envelopes and the foreground clouds, and higher in protostellar outflows. These results are similar to those for low-mass protostars, where outflows dominate the \hho\ line profiles \citep{kristensen:ngc1333}. Envelopes and foreground clouds are barely visible (if at all) in the data for low-mass protostars, which is very likely just due to limited sensitivity.

Several projects are planned to follow up on these initial results.
The \hho\ 557\,GHz line will be studied toward a few positions in infrared dark clouds. These data will clarify the role of water during the earliest stages of high-mass star formation, and also form a link with low-mass pre-stellar cores \citep{caselli:b68}.
A survey of 19 high-mass protostellar objects in the ground state line of p-\hho\ may reveal basic trends of the \hho\ abundance with luminosity, mass, and other physical properties of the sources such as the presence of a hot core and an ultracompact H{\sc II} region.
At the same time, multi-line studies will be valueable to construct radial abundance profiles of \hho, which constrain possible formation and destruction routes for \hho.
Two-dimensional models may be used to derive detailed models of the source structure, and to evaluate the uncertainties of derived \hho\ abundances.
Large-scale (arcmin-sized) maps of \hho\ emission and absorption towards star-forming complexes will constrain the spatial extent of dense gas and its role in clustered star formation.
These maps may also be used to search for \hhop\ emission, which so far only has been seen in the nucleus of Mrk 231 \citep{vdWerf:mrk231}.
Finally, the PACS instrument will be used for imaging of lines of CO, OH, \hho\ and other molecules at far-infrared wavelengths, which will constrain the distribution of high-excitation molecular gas.
Together, these observations will be a significant step towards understanding the physical and chemical processes during high-mass star formation.

\bibliographystyle{aa}
\bibliography{vdtak}

\begin{thebibliography}{9}
\expandafter\ifx\csname natexlab\endcsname\relax\def\natexlab#1{#1}\fi

\bibitem[{{Caselli} {et~al.}(2010){Caselli}, {Keto}, {Pagani}, {Aikawa},
  {Y{\i}ld{\i}z}, {van der Tak}, {Tafalla}, {Bergin}, {Nisini}, {Codella}, {van
  Dishoeck}, {Bachiller}, {Baudry}, {Benedettini}, {Benz}, {Bjerkeli}, {Blake},
  {Bontemps}, {Braine}, {Bruderer}, {Cernicharo}, {Daniel}, {di Giorgio},
  {Dominik}, {Doty}, {Encrenaz}, {Fich}, {Fuente}, {Gaier}, {Giannini},
  {Goicoechea}, {de Graauw}, {Helmich}, {Herczeg}, {Herpin}, {Hogerheijde},
  {Jackson}, {Jacq}, {Javadi}, {Johnstone}, {J{\o}rgensen}, {Kester},
  {Kristensen}, {Laauwen}, {Larsson}, {Lis}, {Liseau}, {Luinge}, {Marseille},
  {McCoey}, {Megej}, {Melnick}, {Neufeld}, {Olberg}, {Parise}, {Pearson},
  {Plume}, {Risacher}, {Santiago-Garc{\'{\i}}a}, {Saraceno}, {Shipman},
  {Siegel}, {van Kempen}, {Visser}, {Wampfler}, \& {Wyrowski}}]{caselli:b68}
{Caselli}, P., {Keto}, E., {Pagani}, L., {et~al.} 2010, \aap, 521, L29

\bibitem[{{Chavarr{\'{\i}}a} {et~al.}(2010){Chavarr{\'{\i}}a}, {Herpin},
  {Jacq}, {Braine}, {Bontemps}, {Baudry}, {Marseille}, {van der Tak},
  {Pietropaoli}, {Wyrowski}, {Shipman}, {Frieswijk}, {van Dishoeck},
  {Cernicharo}, {Bachiller}, {Benedettini}, {Benz}, {Bergin}, {Bjerkeli},
  {Blake}, {Bruderer}, {Caselli}, {Codella}, {Daniel}, {di Giorgio}, {Dominik},
  {Doty}, {Encrenaz}, {Fich}, {Fuente}, {Giannini}, {Goicoechea}, {de Graauw},
  {Hartogh}, {Helmich}, {Herczeg}, {Hogerheijde}, {Johnstone}, {J{\o}rgensen},
  {Kristensen}, {Larsson}, {Lis}, {Liseau}, {McCoey}, {Melnick}, {Nisini},
  {Olberg}, {Parise}, {Pearson}, {Plume}, {Risacher}, {Santiago-Garc{\'{\i}}a},
  {Saraceno}, {Stutzki}, {Szczerba}, {Tafalla}, {Tielens}, {van Kempen},
  {Visser}, {Wampfler}, {Willem}, \& {Y{\i}ld{\i}z}}]{chavarria:w3irs5}
{Chavarr{\'{\i}}a}, L., {Herpin}, F., {Jacq}, T., {et~al.} 2010, \aap, 521, L37

\bibitem[{{Kristensen} {et~al.}(2010){Kristensen}, {Visser}, {van Dishoeck},
  {Y{\i}ld{\i}z}, {Doty}, {Herczeg}, {Liu}, {Parise}, {J{\o}rgensen}, {van
  Kempen}, {Brinch}, {Wampfler}, {Bruderer}, {Benz}, {Hogerheijde}, {Deul},
  {Bachiller}, {Baudry}, {Benedettini}, {Bergin}, {Bjerkeli}, {Blake},
  {Bontemps}, {Braine}, {Caselli}, {Cernicharo}, {Codella}, {Daniel}, {de
  Graauw}, {di Giorgio}, {Dominik}, {Encrenaz}, {Fich}, {Fuente}, {Giannini},
  {Goicoechea}, {Helmich}, {Herpin}, {Jacq}, {Johnstone}, {Kaufman}, {Larsson},
  {Lis}, {Liseau}, {Marseille}, {McCoey}, {Melnick}, {Neufeld}, {Nisini},
  {Olberg}, {Pearson}, {Plume}, {Risacher}, {Santiago-Garc{\'{\i}}a},
  {Saraceno}, {Shipman}, {Tafalla}, {Tielens}, {van der Tak}, {Wyrowski},
  {Beintema}, {de Jonge}, {Dieleman}, {Ossenkopf}, {Roelfsema}, {Stutzki}, \&
  {Whyborn}}]{kristensen:ngc1333}
{Kristensen}, L.~E., {Visser}, R., {van Dishoeck}, E.~F., {et~al.} 2010, \aap,
  521, L30

\bibitem[{{Marseille} {et~al.}(2010){Marseille}, {van der Tak}, {Herpin},
  {Wyrowski}, {Chavarr{\'{\i}}a}, {Pietropaoli}, {Baudry}, {Bontemps},
  {Cernicharo}, {Jacq}, {Frieswijk}, {Shipman}, {van Dishoeck}, {Bachiller},
  {Benedettini}, {Benz}, {Bergin}, {Bjerkeli}, {Blake}, {Braine}, {Bruderer},
  {Caselli}, {Caux}, {Codella}, {Daniel}, {Dieleman}, {di Giorgio}, {Dominik},
  {Doty}, {Encrenaz}, {Fich}, {Fuente}, {Gaier}, {Giannini}, {Goicoechea}, {de
  Graauw}, {Helmich}, {Herczeg}, {Hogerheijde}, {Jackson}, {Javadi}, {Jellema},
  {Johnstone}, {J{\o}rgensen}, {Kester}, {Kristensen}, {Larsson}, {Laauwen},
  {Lis}, {Liseau}, {Luinge}, {McCoey}, {Megej}, {Melnick}, {Neufeld}, {Nisini},
  {Olberg}, {Parise}, {Pearson}, {Plume}, {Risacher}, {Roelfsema},
  {Santiago-Garc{\'{\i}}a}, {Saraceno}, {Siegel}, {Stutzki}, {Tafalla}, {van
  Kempen}, {Visser}, {Wampfler}, \& {Y{\i}ld{\i}z}}]{marseille:envelopes}
{Marseille}, M.~G., {van der Tak}, F.~F.~S., {Herpin}, F., {et~al.} 2010, \aap,
  521, L32

\bibitem[{{Neufeld} {et~al.}(2010){Neufeld}, {Goicoechea}, {Sonnentrucker},
  {Black}, {Pearson}, {Yu}, {Phillips}, {Lis}, {de Luca}, {Herbst}, {Rimmer},
  {Gerin}, {Bell}, {Boulanger}, {Cernicharo}, {Coutens}, {Dartois},
  {Kazmierczak}, {Encrenaz}, {Falgarone}, {Geballe}, {Giesen}, {Godard},
  {Goldsmith}, {Gry}, {Gupta}, {Hennebelle}, {Hily-Blant}, {Joblin},
  {Ko{\l}os}, {Kre{\l}owski}, {Mart{\'{\i}}n-Pintado}, {Menten}, {Monje},
  {Mookerjea}, {Perault}, {Persson}, {Plume}, {Salez}, {Schlemmer}, {Schmidt},
  {Stutzki}, {Teyssier}, {Vastel}, {Cros}, {Klein}, {Lorenzani}, {Philipp},
  {Samoska}, {Shipman}, {Tielens}, {Szczerba}, \& {Zmuidzinas}}]{neufeld:hf}
{Neufeld}, D.~A., {Goicoechea}, J.~R., {Sonnentrucker}, P., {et~al.} 2010,
  \aap, 521, L10

\bibitem[{{Van der Tak} {et~al.}(2010){Van der Tak}, {Marseille}, {Herpin},
  {Wyrowski}, {Baudry}, {Bontemps}, {Braine}, {Doty}, {Frieswijk}, {Melnick},
  {Shipman}, {van Dishoeck}, {Benz}, {Caselli}, {Hogerheijde}, {Johnstone},
  {Liseau}, {Bachiller}, {Benedettini}, {Bergin}, {Bjerkeli}, {Blake},
  {Bruderer}, {Cernicharo}, {Codella}, {Daniel}, {di Giorgio}, {Dominik},
  {Encrenaz}, {Fich}, {Fuente}, {Giannini}, {Goicoechea}, {de Graauw},
  {Helmich}, {Herczeg}, {J{\o}rgensen}, {Kristensen}, {Larsson}, {Lis},
  {McCoey}, {Neufeld}, {Nisini}, {Olberg}, {Parise}, {Pearson}, {Plume},
  {Risacher}, {Santiago}, {Saraceno}, {Tafalla}, {van Kempen}, {Visser},
  {Wampfler}, {Y{\i}ld{\i}z}, {Ravera}, {Roelfsema}, {Siebertz}, \&
  {Teyssier}}]{vdTak:DR21}
{Van der Tak}, F.~F.~S., {Marseille}, M.~G., {Herpin}, F., {et~al.} 2010, \aap,
  518, L107

\bibitem[{{Van der Werf} {et~al.}(2010){Van der Werf}, {Isaak}, {Meijerink},
  {Spaans}, {Rykala}, {Fulton}, {Loenen}, {Walter}, {Wei{\ss}}, {Armus},
  {Fischer}, {Israel}, {Harris}, {Veilleux}, {Henkel}, {Savini}, {Lord},
  {Smith}, {Gonz{\'a}lez-Alfonso}, {Naylor}, {Aalto}, {Charmandaris}, {Dasyra},
  {Evans}, {Gao}, {Greve}, {G{\"u}sten}, {Kramer}, {Mart{\'{\i}}n-Pintado},
  {Mazzarella}, {Papadopoulos}, {Sanders}, {Spinoglio}, {Stacey}, {Vlahakis},
  {Wiedner}, \& {Xilouris}}]{vdWerf:mrk231}
{Van der Werf}, P.~P., {Isaak}, K.~G., {Meijerink}, R., {et~al.} 2010, \aap,
  518, L42

\bibitem[{{Van Dishoeck} {et~al.}(2011){Van Dishoeck}, {Kristensen}, {Benz},
  {a}, {b}, {c}, \& {d}}]{ewine:wish}
{Van Dishoeck}, E., {Kristensen}, L., {Benz}, A., {et~al.} 2011, PASP, in
  press, arxiv:1012.4570

\bibitem[{{Wyrowski} {et~al.}(2010){Wyrowski}, {van der Tak}, {Herpin},
  {Baudry}, {Bontemps}, {Chavarria}, {Frieswijk}, {Jacq}, {Marseille},
  {Shipman}, {van Dishoeck}, {Benz}, {Caselli}, {Hogerheijde}, {Johnstone},
  {Liseau}, {Bachiller}, {Benedettini}, {Bergin}, {Bjerkeli}, {Blake},
  {Braine}, {Bruderer}, {Cernicharo}, {Codella}, {Daniel}, {di Giorgio},
  {Dominik}, {Doty}, {Encrenaz}, {Fich}, {Fuente}, {Giannini}, {Goicoechea},
  {de Graauw}, {Helmich}, {Herczeg}, {J{\o}rgensen}, {Kristensen}, {Larsson},
  {Lis}, {McCoey}, {Melnick}, {Nisini}, {Olberg}, {Parise}, {Pearson}, {Plume},
  {Risacher}, {Santiago}, {Saraceno}, {Tafalla}, {van Kempen}, {Visser},
  {Wampfler}, {Y{\i}ld{\i}z}, {Black}, {Falgarone}, {Gerin}, {Roelfsema},
  {Dieleman}, {Beintema}, {de Jonge}, {Whyborn}, {Stutzki}, \&
  {Ossenkopf}}]{wyrowski:h2o+}
{Wyrowski}, F., {van der Tak}, F., {Herpin}, F., {et~al.} 2010, \aap, 521, L34

\end{thebibliography}

\end{document}